\documentclass[twocolumn]{aastex631}

\shorttitle{Third Epoch Proper Motions in Kepler's SNR}
\shortauthors{Coffin et al.}
\graphicspath{{./}{figures/}}

\begin{document}

\title{A Third Epoch Proper Motion Study of The Forward Shock in Kepler's Supernova Remnant}

\author[0000-0003-3038-80457]{Sadie C. Coffin}
\affiliation{Southeastern Universities Research Association \\
1201 New York Avenue NW,
Suite 430 \\
Washington, DC 20005, USA}
\affiliation{X-ray Astrophysics Laboratory NASA/GSFC \\
Greenbelt, MD 20771}
\affiliation{Center for Research and Exploration in Space Science and Technology \\
Greenbelt, MD 20771}

\author[0000-0003-2063-381X]{Brian J. Williams}
\affiliation{X-ray Astrophysics Laboratory NASA/GSFC \\
Greenbelt, MD 20771}

\author[0000-0002-1104-7205]{Satoru Katsuda}
\affiliation{Graduate School of Science and Engineering, Saitama University, Saitama, Japan}




\begin{abstract}

We present measurements of the expansion of Kepler's Supernova Remnant (SNR) over three epochs of Chandra X-ray observations from 2000, 2006, and 2014. As the remnant of a historical supernova (observed in 1604 CE), Kepler's SNR presents the rare opportunity to study the dynamical evolution of such an object in real time. Measurements of the asymmetry in forward shock velocity can also provide insight into the nature of the explosion and density of the circumstellar material. Combining data from 2014 with previous epochs in 2000 and 2006, we can observe the proper motion of filaments along the outer rim of the SNR. Prior studies of Kepler's SNR have shown proper motion differences up to a factor of 3 between northern and southern regions around the remnant. With the longer time baseline we use here, we find results that are consistent with previous studies, but with smaller uncertainties. Additionally, by adding a third epoch of observations, we search for any systemic change in the velocity in the form of a deceleration of the blast wave, as was recently reported in Tycho's SNR. We find little to no conclusive evidence of such deceleration, and conclude that Kepler's SNR is encountering circumstellar material that is roughly constant in density, though substantially varied around the periphery.

\end{abstract}



\section{Introduction} \label{sec:intro}

Kepler's supernova remnant (SNR; hereafter, Kepler) is one of a small handful of confirmed historical supernovae (SNe), and is definitively associated with the ``stella nova'' observed in 1604 \citep{stephenson02}. First suggested as the remnant of a ``supernova of type I'' by \cite{baade43}, its status as a Type Ia SNR has been effectively confirmed by numerous modern authors using X-ray telescopes such as {\it Ginga}, {\it ASCA}, {\it Chandra}, and {\it XMM-Newton} \citep{hatsukade90,kinugasa99,cassam-chenai04,reynolds07}. The remnant is dominated by the thermal X-ray plasma resulting from heating of gas by both the forward and reverse shocks, but the outer rim is almost entirely nonthermal synchrotron emission \citep{bamba05}. This nonthermal emission was recently discovered to extend up to at least 30 keV \citep{nagayoshi21}.

It has been known for some time that the density of the ambient medium surrounding Kepler must be quite high \citep{hughes85}. The luminosity at all wavelengths is extraordinarily high, implying a dense environment that the forward shock wave is encountering. This is somewhat puzzling, as the remnant has a high galactic latitude (+6.8$^{\circ}$), leading to a height above the galactic plane of 600 pc at a distance of 5.1 kpc as reported in \cite{sankrit16}. 

The most logical explanation for this that the forward shock from the supernova is interacting with a circumstellar medium (CSM) produced by the progenitor system \citep{reynolds07}. \cite{blair07} used {\it Spitzer} infrared imaging observations of Kepler to measure this density, finding post-shock densities (infrared emission from dust grains is only sensitive to the post-shock conditions of the plasma) of $\sim 10-20$ cm$^{-3}$ in the bright northern regions, with densities an order of magnitude lower in the faint southern portions of the remnant, a result that was confirmed by followup {\it Spitzer} spectroscopic observations in \cite{williams12}. The infrared observations are suggestive of an overall density gradient in the CSM in the approximately north-south direction. \cite{katsuda15} find that Kepler's elemental abundances as inferred from X-ray spectroscopy are most consistent with an overluminous Type Ia event, and suggest a connection between remnants like Kepler and ``91T-like'' SNe. \cite{kasuga21} examined {\it XMM-Newton} RGS spectra of the remnant and found that the CSM component has an overall blueshift of up to 500 km s$^{-1}$, consistent with the picture of a ``runaway'' AGB star.

A natural expectation from such a density contrast would be substantial variations in the speed of the forward shock in different locations around the rim as the blast wave encounters regions of steadily increasing density (moving from south to north). Indeed, this is exactly what was observed in previous X-ray proper motion studies using {\it Chandra}. \cite{katsuda08} and \cite{vink08} (hereafter, K08 and V08) both use {\it Chandra} ACIS imaging observations from two epochs (2000 and 2006) to measure the velocity of the blast wave as a function of azimuthal angle around the periphery of the remnant. They find that the velocity varies by a factor of $\sim 2.5-3$, with the lower velocities occurring in the denser northern regions.

A recent discovery in another young Type Ia SNR has prompted us to re-examine the proper motions of the blast wave in Kepler. In Tycho's SNR, \cite{tanaka21} have recently shown evidence for a rapid deceleration of a large section of the blast wave. By examining the proper motions from multiple epochs as a function of azimuthal angle, they show that the shock speeds in the south and southwest portions of the remnant are slowing down at a rate much faster than would be expected from a remnant undergoing a ``typical'' evolution, even in the Sedov phase where R~$\propto$~$t^{0.4}$ (where R is the radius of the remnant and $t$ is time). The deceleration they observe in Tycho is smooth, i.e., it slowly rises and falls, peaking at a position angle of $\sim 200^{\circ}$, and is confirmed across four epochs of observations.

Motivated by this, we have re-measured the proper motions in Kepler, searching for signs of deceleration in this remnant. Our analysis is virtually identical to that done in K08 and V08, as well as that done in \cite{tanaka21} for Tycho's SNR. In this work, we include a {\it third} epoch of observations taken in 2014 with {\it Chandra}. In addition to searching for a robust sign of deceleration between the 2000-2006 and 2006-2014 epochs, the longer overall time baseline of 14 years (as compared to 6) allows us to report the most accurate measurements ever taken of the X-ray emitting shocks in Kepler. 
\section{Observations} \label{sec:observations}

Kepler has been observed by {\it Chandra} multiple times. It was observed for 50 ks in 2000 (PI: Holt), another 50 ks in 2004 (PI: Rudnick), 750 ks in 2006 (PI: Reynolds), 150 ks in 2014 (PI: Borkowski), and another 150 ks in 2016 with the High-Energy Transmission Grating (HETG; PI: Park). For the purposes of this paper, the HETG data are not of use. Also, since the 2004 observations are bracketed by those in 2000 and 2006 (and we seek the longest possible time baselines), we do not use the 2004 observations here either. We followed standard data reduction procedures, using version 4.12.1 of CIAO to reprocess all epochs.

\begin{figure}[ht!]
\plotone{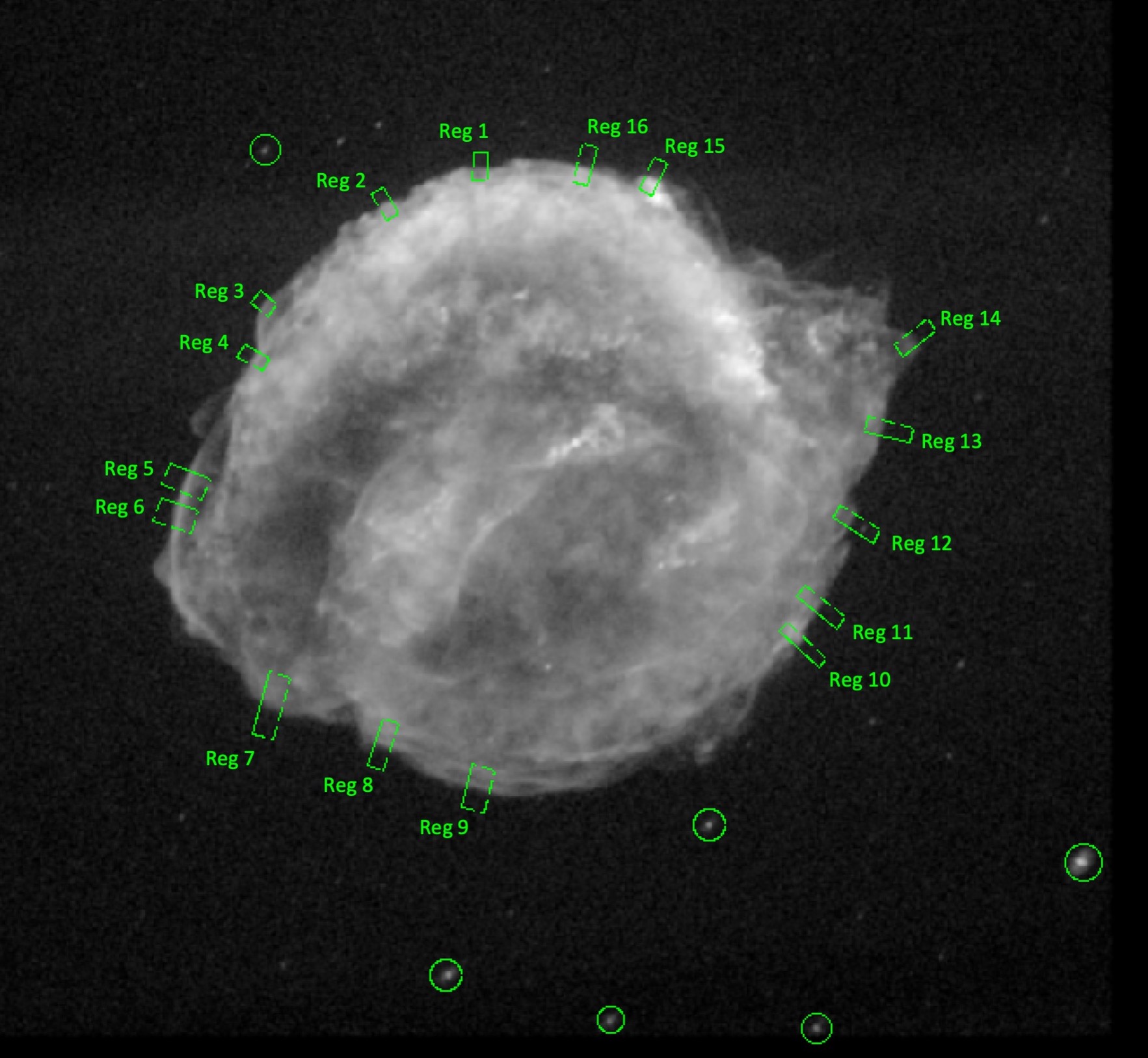}
\caption{Radial profiles were extracted from 16 rectangular regions dispersed around the remnant, labeled in this figure. We chose regions that would be comparable to those examined by K08, save for regions 3, 14 and 15, which we extracted due to interesting features nearby on the remnant but have no comparison in previous studies. The green circles indicate the six sources we used to align the three epochs of data. \label{fig:regions}}
\end{figure}

The astrometric alignment procedure is identical to our previous work on Tycho's SNR \citep{williams16}. We use the CIAO task {\tt wavdetect} to detect point sources in the field. The algorithm ``found'' several dozen sources across the ACIS field, but upon closer inspection, we deemed six of these nearby point sources to be good sources, present at a high signal-to-noise level in all three epochs (ruling out transients) and close enough to the remnant to be minimally affected by {\it Chandra's} degrading spatial resolution off-axis. These six sources are shown in Figure~\ref{fig:regions}. We then used the CIAO tasks {\tt wcs\_match} and {\tt wcs\_update} to reproject the event files to identical wcs coordinates. We use the deep 2006 observations as the relative ``reference'' frame to which all other epochs are aligned. We smooth the X-ray images slightly, using a 2-pixel Gaussian. This has virtually no effect on the profile shapes, but significantly decreases the pixel-to-pixel Poisson noise level.

\section{Proper Motion Measurements} 

We measure the proper motion in all three epochs from a total of 16 regions around the edge of the remnant. These regions were chosen to {\it approximately} correspond to those used in K08, though they are not exactly identical (and need not be, as we are not doing a direct comparison with the results from that paper). All 16 regions, with their exact profile extraction rectangles, are shown in Figure~\ref{fig:regions}. In general, we made the regions as narrow as we could, while still ensuring a high signal-to-noise ratio. We number the regions 1-16, and include the position angle as measured from the approximate geometric center of the remnant. For the purposes of this work, the position angles of the regions themselves are not relevant, but we include them as a reference and for ease of comparison with previous (or future) works. We use the longest time baseline possible: the 2000 and 2014 epochs.

\begin{figure}[ht!]
\plotone{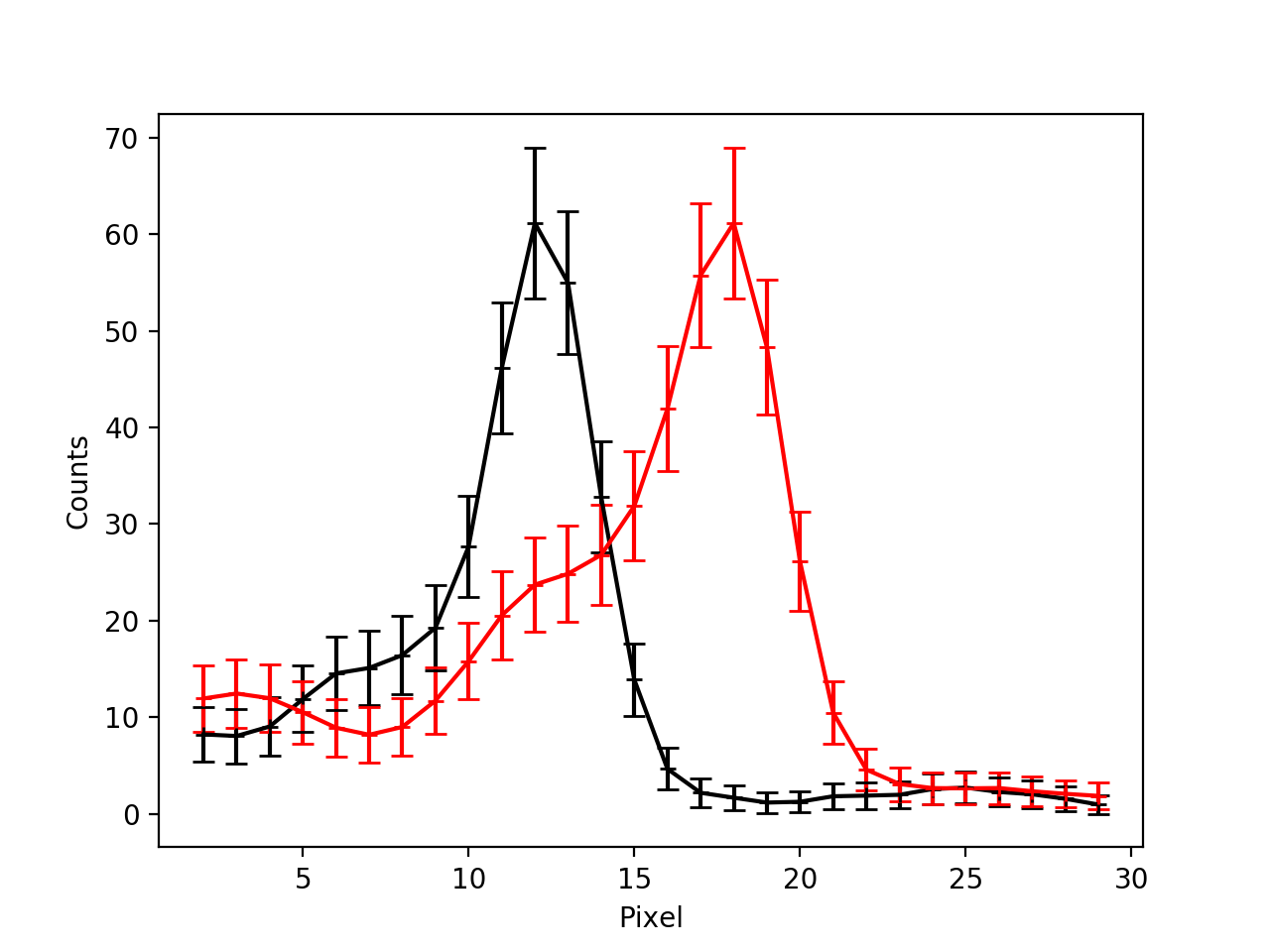}
\caption{This example shows a ``well-behaved'' profile from region 4. \label{fig:good}}
\end{figure}

\begin{figure}[ht!]
\plotone{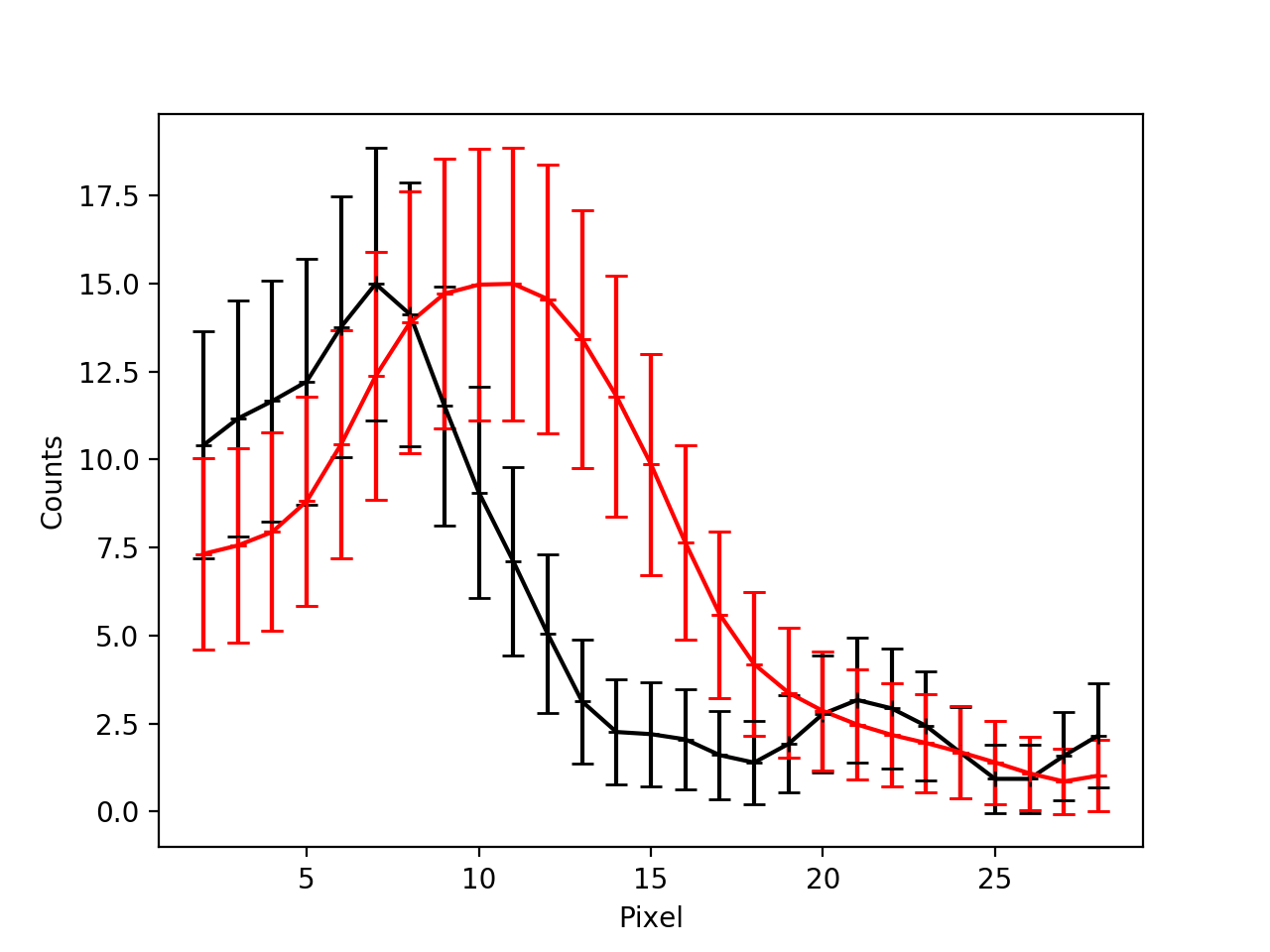}
\caption{This example shows a ``poorly-behaved'' profile from region 14. This is only ``poorly-behaved'' in the relative sense that the slope of the shock front is {\it slightly} different between epochs. \label{fig:bad}}
\end{figure}

We chose regions where the profiles between epochs are generally ``well-behaved,'' meaning that the radial profile has nearly exactly the same shape in all three epochs. An example of a ``well-behaved'' region is shown in Figure~\ref{fig:good}, and the vast majority of our regions fell into this category. For a few regions, such as region 14, the shape of the profile across epochs is very similar, but not quite identical. We call these regions ``poorly-behaved,'' and show an example of such a region in Figure~\ref{fig:bad}. We use the term ``poorly-behaved'' only in a relative sense; in the absolute sense, all 16 regions are quite good, and the $\chi^{2}$ fitting algorithm we use is very robust in finding the optimal value of the shift between epochs, even taking into account the uncertainty sources mentioned in Section~\ref{sec:observations}.

We report both statistical and systematic uncertainties on our values, both of which are derived in an identical fashion to that done in \cite{williams16}. To do the fits and find the uncertainties, we shift one epoch with respect to the other along a fine grid, minimizing the value of the $\chi^{2}$ parameter. Statistical errors are 90\% confidence limits, and correspond to an increase in $\chi^{2}$ of 2.706. To find the velocity vector of the shock front in each region, we search the parameter space of the angle of the projection region with respect to the shock front, making the assumption that the ``correct'' value of the shock velocity corresponds to the maximum value. In general, these regions are roughly perpendicular to the shock front, though not exactly, nor do they exactly backtrack to a common center. This is not unexpected; a remnant as dynamically evolved as Kepler must have at least small scale variations in the expansion of local regions of the shock front as a simple result of having swept up substantial amounts of circumstellar matter. In fact, we caution that it is likely not feasible to extrapolate the exact explosion center of Kepler from the motions of the forward shock after more than 400 years. 

For the systematic uncertainties, we follow the procedure of \cite{williams16} and K08 and use the value of the proper motion at $\pm$ 2$^{\circ}$ in each direction with respect to the maximum value. We also find small variations in the result depending on the choice of fitting window (along the x-axis in Figure~\ref{fig:good} and Figure~\ref{fig:bad}), so we choose six different fitting windows, averaging the result and including the dispersion within these values as part of the systematic uncertainty as well.

\begin{figure}[ht!]
\plotone{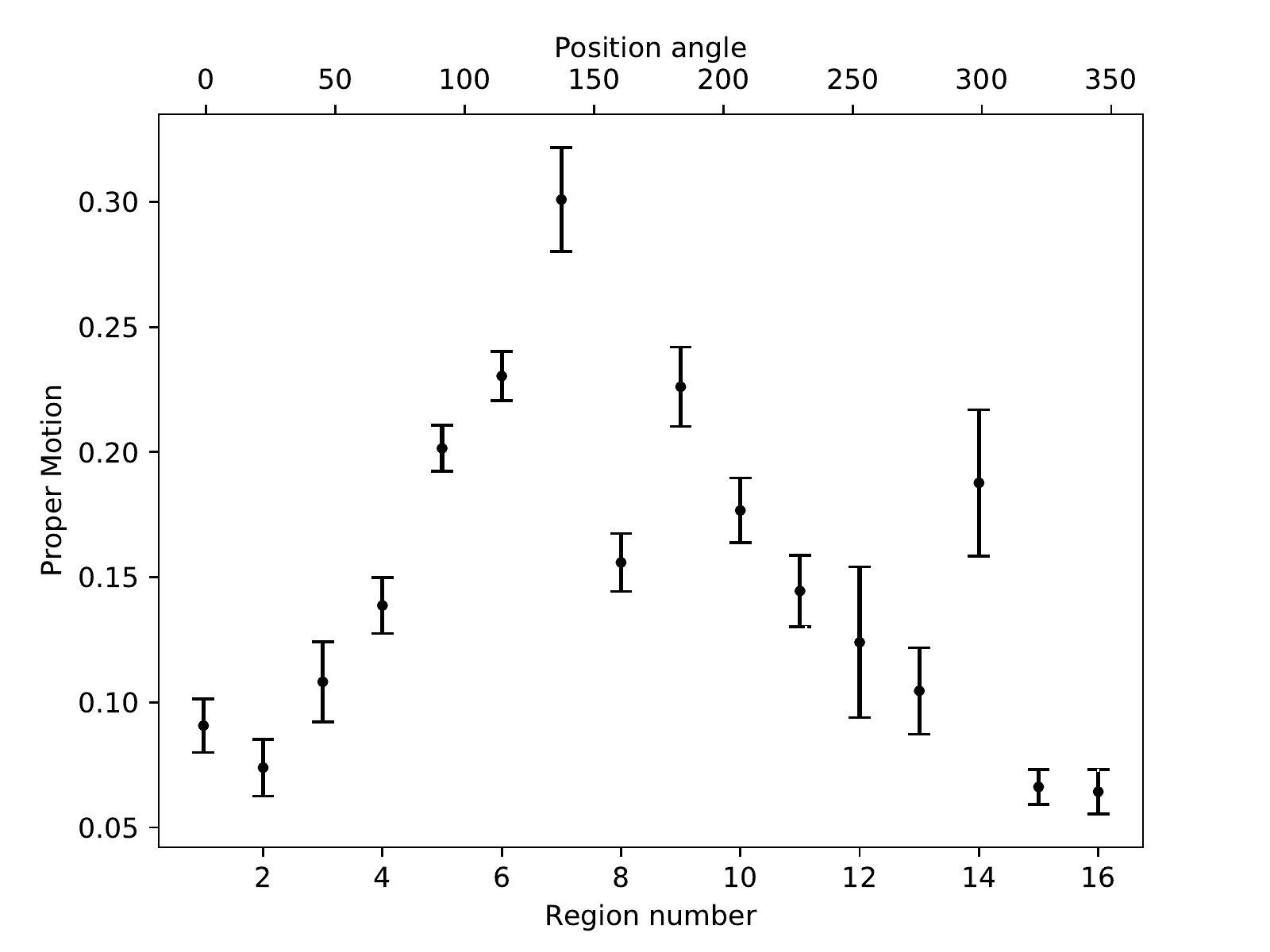}
\caption{The proper motion (in arcsec yr$^{-1}$) within each measured region around the SNR with combined statistical and systematic uncertainty represented in error bars. The trend of faster expansion towards the south and slower expansion along the northern rim aligns with the findings of K08 and V08. The outlier in region 8 was also observed by K08, and the outlier in region 14 is located on a diffuse area on the northwest side of the remnant that was not observed by K08. \label{fig:pm plot}}
\end{figure}

We show our results in Table~\ref{tab:propermotion} and Figure~\ref{fig:pm plot}. We report all proper motion values in this paper in arcsec yr$^{-1}$. If so desired, these values can be scaled to a distance-dependent velocity using the formula 

\begin{equation}
V = 2420(\frac{\theta}{0.1})(\frac{D}{\rm {5.1 kpc}}) \ {\rm {km\ s}^{-1}},
\end{equation}

where $\theta$ is the proper motion in arcsec yr$^{-1}$ and D is the distance to Kepler in kpc. As our results for this paper do not depend on distance, we remain agnostic to the actual distance to Kepler.

\begin{deluxetable*}{cccccc}
\tablenum{1}
\tablecaption{Proper motion measurements and uncertainty values for Kepler's SNR from 2000 to 2014. 
\label{tab:propermotion}}
\tablewidth{0pt}
\tablehead{
\colhead{Region Number} & \colhead{Position Angle} & \colhead{Proper Motion (arcsec yr$^{-1}$)} & \colhead{Syst. Uncertainty} & \colhead{Stat. Uncertainty} & \
}
\startdata
1	&	9	&   0.091	&	0.0002	&	0.0105 \\
2	&	29 &   0.074	&	0.0030	&	0.0083 \\
3	&	58  &   0.108	&	0.0015	&	0.0146 \\
4	&	68  &   0.139	&	0.0035	&	0.0077 \\
5	&	92 &   0.202	&	0.0042	&	0.0050 \\
6	&	97  &   0.230	&	0.0048	&	0.0050 \\
7	&	130  &   0.301	&	0.0043	&	0.0165 \\
8	&	151	&   0.156	&	0.0029	&	0.0085 \\
9	&	171  &   0.226	&	0.0044	&	0.0116 \\
10	&	238 &   0.177	&	0.0029	&	0.0099 \\
11	&	245 &   0.144	&	0.0012	&	0.0131 \\
12	&	262 &   0.124	&	0.0044	&	0.0256 \\
13	&	277 &   0.105	&	0.0027	&	0.0147 \\
14	&	289 &   0.188	&	0.0022	&	0.0270 \\
15	&	336 &   0.066	&	0.0028	&	0.0041 \\
16	&	349  &   0.064	&	0.0026	&	0.0063 \\
\enddata
\tablecomments{Position angle is measured from the approximate geometric center of the remnant. Statistical uncertainty represents an average of the uncertainties above and below the measured value; we did not find a statistically significant difference between the two. Systematic uncertainty is described in the text.}
\end{deluxetable*}

We confirm the results of an overall substantial dependence of velocity on position angle, with much faster shocks in the south. The shocks in the southern portion of the remnant are about three times faster than those along the brighter northern rim. Our results generally agree quite well with those of K08 and V08, even down to the peculiar case of region 8 (which approximately corresponds to ``region 7'' in K08 and position angle 175 in V08). This region has only about half the velocity that the regions on either side of it have. An examination of Figure~\ref{fig:regions} shows that region 8 falls on a section of the remnant that is slightly indented with respect to regions 7 and 9, so this lower proper motion value there is likely real.

\section{Discussion}

The values for the shock velocity that we measure are consistent with the density gradient reported in Kepler. Taking the simple case of ram pressure equilibrium and assuming that $\rho v^{2}$ is constant, the order of magnitude difference in density from the northern edge to the southern edge implies roughly a factor of three difference in shock velocity, which is exactly what is seen in the proper motion measurements. 

It is also interesting to compare our values with those measured from optical {\it HST} data. It is only the northern filaments in Kepler that emit H$\alpha$ emission (plus a few central filaments); this is again consistent with the higher densities in the north. \cite{sankrit16} used two {\it HST} epochs from 2003 and 2013 to measure the velocities along the main shock of the northern rim, finding an average shock velocity of 1690 km s$^{-1}$ (scaled to 5.1 kpc). This corresponds to regions 15, 16, 1, and 2 in our work, and if we average these velocities and scale them to 5.1 kpc using equation 1, we get an average velocity of 1780 km s$^{-1}$, a near perfect agreement and well within uncertainties on both values. This also confirms that both the H$\alpha$ and the nonthermal X-ray synchrotron emission are both being produced at the very edge of the shock front.

Given the \cite{tanaka21} results of the deceleration in Tycho, we re-did all of our measurements using the middle 2006 epoch, thus creating two values for the velocity for all 16 regions: a 2000-2006 epoch, and a 2006-2014 epoch. Following the same procedure as \cite{tanaka21}, we simply compare the results from the two epochs and look for substantial changes in velocity. We note here that deceleration in all SNRs is a given as the shock slowly sweeps up more and more material. However, in most remnants, the deceleration happens over extremely long timescales. Even for remnants in the Sedov phase, where R $\propto$ t$^{0.4}$, the timescales necessary to measure substantial deceleration at {\it Chandra's} resolution are quite long, and the deceleration signal one would hope to measure from this ``natural'' deceleration would be completely swamped by the uncertainties in the measurements.

\begin{figure}[ht!]
\plotone{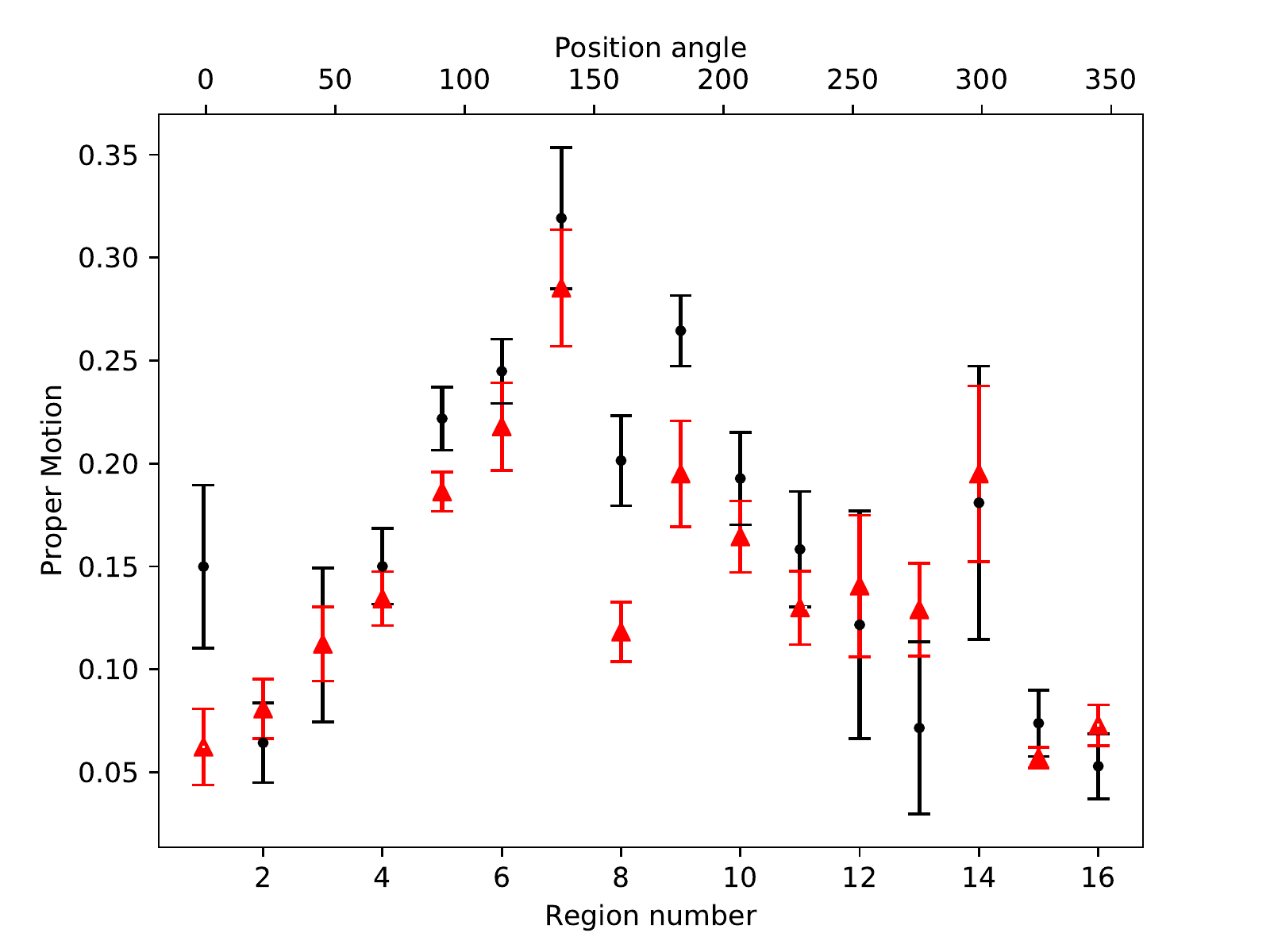}
\caption{Comparing the proper motion (in arcsec yr$^{-1}$) within each region between 2000 and 2006 (black data points) with the proper motion between 2006 and 2014 (red data points) can help reveal any velocity change in the remnant's expansion. For 12 of the 16 regions, we did not find a statistically significant difference in proper motion between these epochs. Regions 1, 5, 8, and 9, dispersed on different sides of the remnant, are the only regions with a statistically significant decrease in proper motion. \label{fig:ab bc plot}}
\end{figure}

What we search for here, as was seen in Tycho, is deceleration that is not possible via normal remnant evolution. \cite{tanaka21} report that the decelerations observed there require a recent encounter with a dense wall of material with a density contrast of 100-1000 higher than inside the wall. The results are shown in Figure~\ref{fig:ab bc plot}. For 12 of the 16 regions, we find no statistically significant shift at all. Most notably, unlike in Tycho, we see no systemic shift in several regions that are spatially correlated to be near each other.

There is a possible hint of deceleration in regions 8 and 9. These two regions are near each other, and both show a small but statistically significant drop in their velocity. Given that region 8 is also the region in which the shock velocity is substantially slower than the surrounding regions (as confirmed by K08 and V08), it is possible that the blast wave in these regions is encountering a localized dense clump of interstellar material. We have examined these two regions in the context of other works that studied Kepler on similar scales, and find nothing particularly unusual about these regions, aside from that mentioned above. Recently, \cite{reynolds21} examined regions very similar to ours, examining the nonthermal properties in detail. They used X-ray and IR observations to infer the efficiencies of magnetic field amplification and electron acceleration at various points around the periphery of Kepler. Our regions 8 and 9 are very similar to their regions 7 and 7b. While there is a large spread in the properties of the nonthermal emission as reported in \cite{reynolds21}, there is nothing particularly peculiar about these two regions.

It is possible that measurement uncertainties beyond what we have accounted for here could be responsible for this apparent drop in velocity. However, we can (at least at this epoch) rule out a large cavity wall that a large section of the blast wave is encountering, as is seen in Tycho. Additional observations of Kepler in future epochs would be useful to further quantify the possible presence of a decelerating blast wave at these and other locations.

\section{Conclusion}

In light of the recent discovery of a rapidly decelerating portion of the shock front in Tycho's SNR, we have re-examined the proper motion in Kepler's SNR based on a third epoch of {\it Chandra} observations in 2014. We have increased the overall baseline for proper motion measurements over previous works by more than a factor of two (6 years vs. 14 years), resulting in the most accurate proper motion measurements of the shock front to date. Our results are consistent with previous works of K08 and V08, and we confirm previous results of shock velocity differences of a factor of $\sim 3$ from the northern to the southern rim, where such differences result from substantially varying densities of the surrounding CSM.

We find little evidence for the relatively large scale decelerations that are seen in Tycho. There are two consecutive regions in the southern portion of the remnant, separated by a few degrees, that do show a statistically significant slowing of the shock front. One of these regions, region 8, is peculiar in being much slower than the surrounding regions, a finding also seen by K08 and V08. It is likely that the shock there is encountering a localized dense clump of material. This region and others should continue to be monitored periodically by future high resolution X-ray observations to search for any additional changes in the velocity.

\begin{acknowledgments}

We thank the anonymous referee for suggestions that improved the quality and clarity of presentation of this manuscript.

\end{acknowledgments}

%




\bibliography{kepler.bib}{}
\bibliographystyle{aasjournal}


\end{document}